\documentclass{emulateapj}
\usepackage{amsmath}
\shorttitle{Example of Draft}
\shortauthors{Sasaki, Asano \& Terasawa}

\begin{document}

\title{
Time-Dependent Stochastic Acceleration Model for the Fermi Bubbles
}
\author{Kento Sasaki, Katsuaki Asano and Toshio Terasawa}
\affil{Institute for Cosmic Ray Research, The University of Tokyo,
5-1-5 Kashiwanoha, Kashiwa, Chiba 277-8582, Japan}
\email{kentos@icrr.u-tokyo.ac.jp, asanok@icrr.u-tokyo.ac.jp, terasawa@icrr.u-tokyo.ac.jp}

%%\date{Submitted; accepted}

\begin{abstract}
We study stochastic acceleration models for the Fermi bubbles.
Turbulence is excited just behind the shock front
via Kelvin--Helmholtz, Rayleigh--Taylor, or Richtmyer--Meshkov instabilities,
and plasma particles are continuously accelerated by the interaction
with the turbulence.
The turbulence gradually decays as it goes away from the shock fronts.
Adopting a phenomenological model for the stochastic acceleration,
we explicitly solve the temporal evolution of the particle energy distribution
in the turbulence.
Our results show that the spatial distribution of high-energy particles
is different from those for a steady solution.
We also show that the contribution of
electrons that escaped from the acceleration regions
significantly softens the photon spectrum.
The photon spectrum and surface brightness profile
are reproduced by our models.
If the escape efficiency is very high,
the radio flux from the escaped low-energy electrons
can be comparable to that of the WMAP haze.
We also demonstrate hadronic models with the stochastic acceleration,
but they are unlikely in the viewpoint of the energy budget.
\end{abstract}


\section{Introduction}
\label{sec:intro}
Gamma-ray data of the {\it Fermi} Large Area Telescope (LAT)
reveal bilobal giant bubbles extending
up to $\sim50^{\circ}$ above and below the Galactic disk,
called the ``Fermi Bubbles (FBs)'' \citep[][]{su10,su12,ack14,yan14}.
There the microwave bubbles also exists in the same region, the WMAP haze \citep{fin04,dob08}.
These huge structures may suggest past large-scaled activities in the Galactic Center,
such as active galactic nucleus (AGN) jet activity \citep[e.g.][]{guo12,yan12},
non-relativistic outflow from the nucleus \citep[e.g.][]{zub11,zub12,mou14},
or wind driven by supernovae \citep[e.g.][]{cro12,lac14}. 
The gamma-ray spectra of the FBs are harder than the ambient diffuse spectrum
of the Galactic halo,
and the surface brightness profiles show a sharp rise at the bubble edges.
The intensity inside the FBs is almost constant,
which requires inhomogeneous gamma-ray emissivity inside the FBs.
If the volume emissivity is constant,
the surface brightness should show a bump-like profile
with gradually rising edges as a result of the projection effect \citep{mer11}.
On the other hand, a localized emissivity at the shock fronts
should yield limb-brightened profiles.
Significantly thick shells are therefore preferable for the emission regions.

As emission mechanisms, hadronic \citep{cro11,fuj13,che15}
and leptonic \citep{che11,che14,che15b,mer11} models were proposed.
In hadronic models,
shock-accelerated protons produce pions via $pp$-collision,
and gamma-rays are emitted from $\pi^0$-decay.
The present Mach number of the shocks is not so large \citep{tah15}
that the direct shock acceleration may not be effective.
High-energy protons that were shock-accelerated past early-stages
may still be confined in the FBs.
Such protons widely distributed in the downstream
may be responsible for the gamma-ray emission
\citep{fuj13,fuj14}.
In leptonic models, high-energy electrons emit gamma-rays
via inverse Compton (IC) scattering.
The short cooling timescale for electrons requires continuous acceleration
in the downstream
to secure a large volume of the emission region.
The second order Fermi acceleration is the most promising candidate
for such an acceleration mechanism.
Electrons may be continuously accelerated by scattering with
turbulences in the FBs.
The stochastic acceleration is a candidate of the particle
acceleration mechanisms in supernova remnants \citep{fan10},
lobes of radio galaxies \citep{har09,osu09},
blazars \citep{asa14,asa15,kak15}, and gamma-ray bursts \citep{asa09}.
\citet[][hereafter MS11]{mer11}
succeeded in reproducing the gamma-ray spectrum
and surface brightness profile with such a stochastic acceleration model,
taking into account the gradual decay of the turbulences in the downstream.
\citet{che14,che15b} also considered the stochastic acceleration
in certain localized regions.

In this paper, we revisit the stochastic acceleration model for the FBs
focusing on the effects of the time-dependence and escape from
the acceleration regions.
We phenomenologically simulate the acceleration
and photon emission of the electrons advected away from the shock front
with evolving acceleration efficiency.
We also show that the emission from
the escaped electrons from the acceleration regions
is important.
Note that MS11 adopted steady solutions for the electron energy distribution,
while the diffusion coefficient in the momentum space is assumed to decay
with time.
In addition, the steady solutions accompany electron escape,
but the escaped electrons are neglected in MS11.
The model in \citet{che15b} is a different type of model, in which
the shock fronts do not play an important role.
Our goal is to find the best phenomenological description of
the diffusion coefficient evolving in the downstream region rather than
probe details of the acceleration mechanism.
The validity of the obtained requirements
for particle acceleration will be testified by future studies.

The radio emission from the WMAP haze is hard to reconcile
both the hadronic and leptonic models.
The amount of the secondary positrons/electrons generated from $\pi^{\pm}$
is insufficient to reproduce the WMAP haze by synchrotron radiation.
Another leptonic component is required to reproduce the WMAP haze
in the hadronic models \citep{fuj14,che15}.
Even for the leptonic models,
a stronger magnetic field than the typical galactic value
($B \sim 4 \mu$ G) is needed in MS11.
We therefore try to search a condition to reproduce
both the FBs and WMAP haze in our picture.

In \S \ref{sec:form}, we present our model assumption and
computing method.
The results are summarized in Sections \ref{sec:lep} and \ref{sec:had}
for leptonic and hadronic models, respectively.
Section \ref{sec:sum} provides a summary and discussion.

\section{Model and Method}
\label{sec:form}

\begin{figure}[!ht]
\centering
\epsscale{1.0}
\plotone{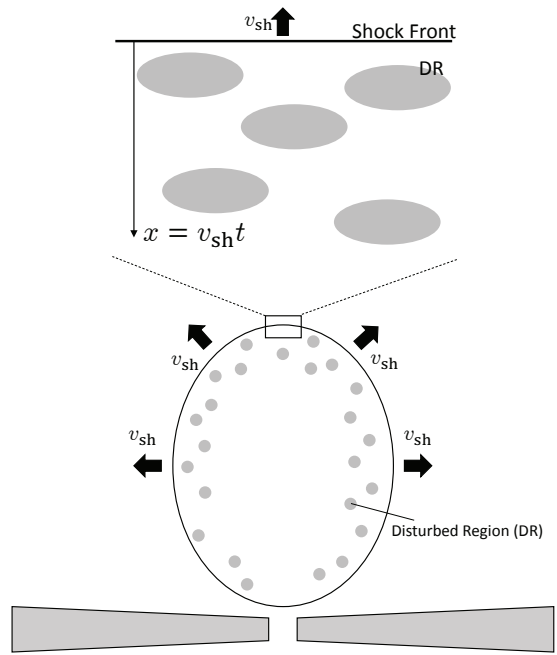}
\caption{Schematic picture of our model.
Plasma particles in the disturbed region,
whose typical size is $L$, are accelerated by turbulences.
The distance from the shock front $x$ and elapsed time $t$
are connected as $x=v_{\rm sh} t$.}
\label{Schpic}
\end{figure}

There may exist a shock front, propagating outward, near the edge of the FBs.
In the downstream of the shock, plasma turbulences are probably excited
via Kelvin--Helmholtz instability \citep{guo12},
Rayleigh--Taylor instability \citep{mue91,bau13,yan13},
or Richtmyer--Meshkov instability \citep{ino09}.
In our model, as assumed in MS11,
electrons/protons are stochastically accelerated
by scattering with the turbulences.
We primarily consider the electron acceleration model,
but hadronic models, where protons are accelerated by turbulences,
are also discussed in this paper.
The particle escape from the acceleration process
is a nontrivial problem.
When the diffusion length scale becomes larger than the size of the FB,
particles can escape from the FBs \citep{ohi11}.
The large size of the FBs and small spatial diffusion coefficient
implied from the efficient scattering in our model suggest that
we can neglect the escape of the particles from the FBs.
Actually, the one-dimensional (1D) models taking into account the diffusion of particles
in \citet{fuj13,fuj14} almost confine the accelerated
particles within the FBs.
As will be shown below, however, the continuous acceleration
models without the escape effect are difficult to reconcile with
the observed spectra.
MS11 considered the steady-state model with the escape effect.
Similarly to their model, we assume patchy disturbed regions
(hereafter DRs)
with a scale of $L$ (see Figure \ref{Schpic}), which may correspond to the size
of the inhomogeneity in the upstream fluid
for the Richtmyer--Meshkov instability model.
In our model, particles are accelerated by the turbulences in the DRs,
and escape with a certain probability.
The DRs are advected away from the shock front
with a speed of $v_{\rm sh}$ as depicted in Figure \ref{Schpic}.

In this case, the evolution of the particle energy distribution
in the DR rest frame is written by the Fokker--Planck equation
\citep[see, e.g.][]{sta08},
%%%%%%%%%%%%%%%%%%%%%%%
\begin{equation}
\frac{\partial n}{\partial{t}}
-\frac{\partial}{\partial p}\left(p^2D_{pp}\frac{\partial}{\partial p}\frac{n}{p^2} \right)
+ \frac{n}{t_{\rm esc}}+\frac{\partial}{\partial p}\left( \frac{dp}{dt}n\right)- Q_{\rm inj}= 0
\label{FPeq}
\end{equation}
%%%%%%%%%%%%%%%%%%%%%%%
where the second--fifth terms represent the energy change
via the stochastic process,
the escape from the DRs,
the cooling by the synchrotron and IC radiation,
and the particle injection, respectively.
The particle density $n(p,t)$ is the average density,
while the actual density is higher according to the inverse of the filling factor
of the DRs.
At the present stage, the filling factor is unpredictable,
depending on the mechanism of the hydrodynamical instability
or the density-inhomogeneity formation in the upstream.
We neglect the volume change of the DR so that the adiabatic cooling term
is omitted.

The properties of the turbulences responsible for the particle acceleration
are highly uncertain;
the initial ratio of the average Alfv\'en speed
to the turbulent velocity, the dominant wave mode
\citep[Alfv\'en or acoustic etc., see, e.g.][]{cho06},
and the range of the wavenumber.
The resonant interaction between particles
and the Alfv\'en wave \citep[see, e.g.][]{bla87,sta08}
is a possible candidate for the electron acceleration mechanism.
Alternatively, fast mode waves have been frequently considered
to be the scatterers of electrons \citep[e.g.][]{liu06,fan10},
though the cut-off length scale due to wave damping
depends on the unknown initial condition.
Moreover, the recent simulations \citep{lyn14}
showed that slow mode waves can be dominant accelerators as well,
even when the wave-particle resonance condition is not satisfied.
It may be difficult to specify the acceleration process
in the FBs from the observations at the present stage.
Even if we successfully reproduce the FBs by a detailed model of the temporal evolution
motivated by a specific physical mechanism,
this may not necessarily mean the validity of the supposed mechanism.
Therefore, we first find a phenomenological evolution
of the acceleration process to reproduce the FBs.
Then, the required evolution should be examined by future plasma simulations.

We treat only ultra-relativistic particles,
then the momentum diffusion coefficient can be phenomenologically described as
%%%%%%%%%%%%%%%%%%%%%%%
\begin{equation}
D_{pp}(p,t) \equiv \frac{2}{3}\beta_{\rm W}^2 D p^{q},
\label{Dpp}
\end{equation}
%%%%%%%%%%%%%%%%%%%%%%%
irrespective of the wave mode \citep[e.g.][]{bla87,cho06}.
Here, $v_{\rm W}=\beta_{\rm W}c$ is the effective wave velocity,
which may correspond to the sound velocity or the Alfv\'en velocity.
The factor $\beta_{\rm W}^2$ implies that the average energy gain per scattering
is proportional to $\beta_{\rm W}^2 p$.
The coefficient $D$ is defined with the mean free path $l_{\rm mfp}(p)$
as $D \equiv p^{2-q}c/l_{\rm mfp}$.
In this paper, we consider a case of $q=2$ only (hard sphere approximation)
to produce significantly soft gamma-ray spectra for the sub-GeV region
to be consistent with the observations.
In this case, $l_{\rm mfp}$ and the acceleration timescale $\sim p^2/D_{pp}$
become energy-independent.
The same assumption $q=2$ as that in MS11 may be consistent with the nonresonant scattering
by acoustic modes in turbulence with a typical eddy size \citep{ptu88,cho06,lyn14}.
Even for resonant scattering by the Alfv\'en wave,
the wavenumber spectrum $\propto k^{-2}$
in simulations of freely decaying magnetohydrodynamic turbulence
\citep{chr01,bra15} may support the index $q=2$.

The spatial diffusion coefficient $D_{xx}\sim (1/3)c l_{\rm mfp}$
is also described with $D$ as
%%%%%%%%%%%%%%%%%%%%%%%
\begin{equation}
D_{xx}(t) \equiv \frac{c^2}{3D}.
\label{Dxx}
\end{equation}
%%%%%%%%%%%%%%%%%%%%%%%
While MS11 adopted a model with decaying fast magnetosonic waves
based on \citet{fan10}, we simply assume a power-law temporal evolution for $D$.
Setting $t=0$ at the shock front, we express
%%%%%%%%%%%%%%%%%%%%%%%
\begin{eqnarray}
D(t) &\equiv& D_0 \left(1+\frac{t}{t_0}\right)^{-\alpha},
\end{eqnarray}
%%%%%%%%%%%%%%%%%%%%%%%
where the typical crossing timescale $t_0 \equiv L/v_{\rm sh}$
is introduced to avoid divergence at $t=0$.
The escape timescale from the DRs is written as $t_{\rm esc} \equiv L^2/D_{xx}$.

The decay index $\alpha$ is a free parameter in this paper.
However, as will be shown, $\alpha \sim 1$ seems favorable for the FBs.
Interestingly, three-dimensional (3D)
numerical simulations of freely decaying magnetohydrodynamic turbulences
show inverse cascades, in which the energy density of the turbulence
decays ($U \propto t^{-0.7}$--$t^{-1.1}$)
with a wavenumber spectrum $\propto k^{-2}$, while the eddy size grows
($l_{\rm edd} \propto t^{0.4}$--$t^{0.5}$) \citep{chr01,bra15}.
For both the resonant scattering with the Alfv\'en wave \citep{bla87}
and the nonresonant scattering with acoustic waves \citep{lyn14},
$D \propto U/l_{\rm edd}$
so that $\alpha \sim 1$ may be acceptable.
In any case, note that our purpose in this paper
is to constrain the phenomenological parameters
such as $D_0$ rather than determine the turbulence decay process
and the microscopic mechanism of the electron acceleration.

\begin{figure}[!htb]
\centering
\epsscale{1.0}
\plotone{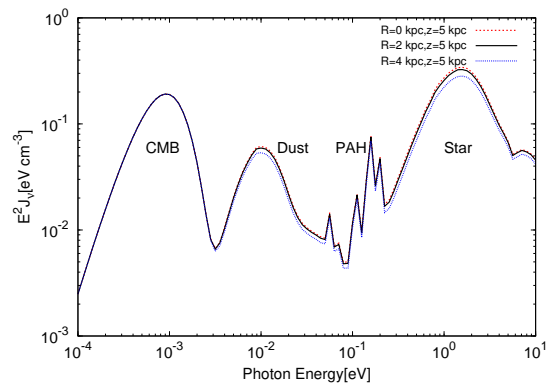}
\caption{Interstellar radiation field from GALPROP v54.
We adopt the black solid line ($R=2$ kpc, $z=5$ kpc)
as a representative field of the whole region.}
\label{bgph}
\end{figure}

The synchrotron and IC coolings for electrons are numerically treated
with the standard method including the Klein--Nishina effect.
The magnetic field is assumed as $B_0=4 \mu\mbox{G}$ for the entire region.
The model of the interstellar radiation field is taken
from the GALPROP code\footnote{http://galprop.stanford.edu/}
\citep[v54,][and references therein]{vla11},
in which the model of \citet{GALPROP} is adopted.
We do not take into account the inhomogeneity of the radiation field,
but adopt a representative field at $R=2$ kpc and $z=5$ kpc in the
Galactocentric coordinate as shown in Figure \ref{bgph}
(the energy density $U_{\rm ph}=1.5 \times 10^{-12}~\mbox{erg}~\mbox{cm}^{-3}$).
We neglect the cooling effect for protons.

The electron injection into the stochastic acceleration
is also an unknown process similarly to that in the shock acceleration.
The background cosmic-rays distributed in the upstream region in advance
may be re-accelerated by the turbulence (see \S \ref{sec:sum}).
Here, we simply assume the injection term
with the same power-law temporal evolution as $D$:
%%%%%%%%%%%%%%%%%%%%%%%
\begin{equation}
Q_{\rm inj}(p,t) = Q_0 \delta(p-p_0) \left(1+\frac{t}{t_0}\right)^{-\alpha}.
\end{equation}
%%%%%%%%%%%%%%%%%%%%%%%
An injection rate proportional to the inverse of the acceleration timescale
may be reasonable as a first-step assumption.
The injection energies are taken as $c p_0=10^8$ eV and $10^{10}$ eV for electrons and protons,
respectively.
Though the actual injection energy may be lower than our assumptions,
note that the parameter $Q_0$ is related to the electron flux
at $p=p_0$ in momentum space for electrons accelerated from lower momenta.

While MS11 counted electrons only in the DRs,
the escaped particles should contribute to the emissions of the FBs as well.
The escaped particles cool without the acceleration effect.
We calculate the evolution of the escaped particles as
%%%%%%%%%%%%%%%%%%%%%%%
\begin{equation}
\frac{\partial n_{\rm esc}}{\partial{t}}+\frac{\partial}{\partial p}
\left( \frac{dp}{dt} n_{\rm esc} \right)- Q^{\rm esc}_{\rm inj}= 0,
\label{FPesc}
\end{equation}
%%%%%%%%%%%%%%%%%%%%%%%
where the injection term for the escaped particles $Q^{\rm esc}_{\rm inj}$
is equal to $n/t_{\rm esc}$.
For simplicity, we neglect the reentry of the escaped particles into the DRs.

We simultaneously solve equations (\ref{FPeq}) and (\ref{FPesc}).
In our method, the obtained temporal evolutions for $n(t)$ and $n_{\rm esc}(t)$
are interpreted as the spatial distributions of $n(x)$ and $n_{\rm esc}(x)$
at the present moment.
The time $t$ and the distance from the shock front $x$
are connected as $x=v_{\rm sh} t$ (see Figure \ref{Schpic}), then the radial position
is written as $R=R_{\rm sh}-x$, where $R_{\rm sh}$ is the radius of the shock front.
Hereafter, we adopt a constant value of $v_{\rm sh}=250~\mbox{km}~\mbox{s}^{-1}$,
which is consistent with the X-ray observations of the FBs \citep{tah15}.
Of course, the expansion speed of the FBs may be not constant and depend on
the energy injection history and the density profile
\citep[e.g.][]{guo12,zub12,bau13,fuj13,lac14}.
The deceleration/acceleration of the shock front makes our model complicated,
and forces us to (at least) a 1D calculation rather than a one-zone
approximation.
To simplify our model, we assume that the shock speed has not changed drastically
in the last period of a few megayears.
This assumption is consistent with almost constant shock velocity
at the last stage
in the prompt energy injection model of \citet{fuj14}
based on the galactic halo profile in \citet{guo12}.
Even in the long energy injection model of \citet{cro15}, the shock velocity
is almost constant after a few megayears from the onset of the energy injection.
The almost constant electron density at a scale of 1--2 kpc behind the shock
in the simulations of $\sim 10$ Myr activity of the galactic nucleus
by \citet{mou14} also supports our simple assumption indirectly.

We numerically calculate the spectral evolutions of $n$ and $n_{\rm esc}$
as far as $x=2$ kpc ($0<t<7.8$ Myr).
This provides us with the radial dependence of the emissivity.
Assuming spherical symmetry, we calculate the surface brightness
by integrating the emissivity along the line of sight.
We neglect the emissions from the region inside $R_{\rm sh}-2$ kpc.
The emission processes we consider are synchrotron, IC,
and pion decay ($\pi^0 \to 2 \gamma$) arising from a $pp$ collision.
The numerical method to calculate such emission spectra is the same as
that in \citet{asa12} and \citet{mur12}.

If the plasmas in the FBs have temperatures of $\sim 0.3$ keV \citep{tah15},
density $\sim 10^{-3}~\mbox{cm}^{-3}$, and magnetic field $\sim 4 \mu\mbox{G}$,
both the Alfv\'en speed and sound speed are on the order of
$10^{7}~\mbox{cm}~\mbox{s}^{-1}$.
So we adopt $\beta_{\rm W}=5.0\times 10^{-4}$ hereafter.
Even if $\beta_{\rm W}$ is different, we can obtain similar results
by scaling the parameters $D$ and $L$.

%%%%%%%%%%%%%%%%%%%%%%%

\section{Leptonic Models}
\label{sec:lep}

The phenomenological model parameters are $B_0$, $D_0$, $\alpha$, $Q_0$,
$L$, and $R_{\rm sh}$.
The value $D_0$ adjusts the acceleration efficiency,
$Q_0$ normalizes the flux level, $L$ controls
the timescales $t_0$ and $t_{\rm esc}$,
and $R_{\rm sh}$ is the actual shock front that
may be displaced from the observed edge of the FBs.
For leptonic models, we show three models in which
the temporal decay indices are assumed as $\alpha=1$.
The other parameters are summarized in Table \ref{para_lep}.

%%%%%%%%%%%%%%%%%%%%%%%
\begin{deluxetable*}{lccccc}
\tablewidth{0pt}
\tablecaption{Parameters for leptonic models \label{para_lep}}
\tablehead{
\colhead{Model}                    & \colhead{$L$ (pc)}                               &
\colhead{$D_0$ ($\mbox{s}^{-1}$)}  & \colhead{$Q_0$ ($\mbox{cm}^{-3}~\mbox{s}^{-1}$)} &
\colhead{$R_{\rm sh}$ (kpc)}       & \colhead{$E_{\rm tot}$ (erg)} 
}
\startdata
NoE (No Escape) & 1500\tablenotemark{a} & $1.5 \times 10^{-7}$ &
$5.9 \times 10^{-30}$ & 4.4 & $1.1 \times 10^{53}$ \\
MiE (Mild Escape) & 180 & $4.0 \times 10^{-7}$ &
$4.2 \times 10^{-29}$ & 3.6 & $ 2.6 \times10^{53}$ \\
EfE (Efficient Escape) & 10 & $4.0 \times 10^{-6}$ &
$7.4 \times10^{-27}$ & 3.0 & $2.0 \times 10^{54}$
\enddata
\tablenotetext{a}{In model NoE, we neglect the escape effect,
so $L$ provides the timescale $t_0$ only.}
\end{deluxetable*}
%%%%%%%%%%%%%%%%%%%%%%%

As a test case, we adopt the exact same values and evolutions for $D_{pp}$ and $t_{\rm esc}$
as those in MS11, in which fast magnetosonic waves
injected at a relatively large scale of 2 kpc
are responsible for the acceleration.
While MS11 assumed the steady-state solution for each radius,
we numerically follow the temporal evolutions of the electron energy distribution
with the evolving $D_{pp}$ and constant injection rate $Q_{\rm inj}$.
\footnote{MS11 did not explicitly declare the injection rate profile.}
Although the obtained results are different from the steady-state solutions,
we omit to show those results in this paper.
The parameters for model NoE, whose results will be shown in the next subsection,
are adjusted to yield a similar $D_{pp}$ evolution to the model of MS11.
\footnote{The evolution of $D_{pp}$ is not a power-law form
of $x$ in MS11.}
Consequently, the obtained spectra for electrons and gamma-rays also become similar
to those in MS11.

\subsection{No Escape Model}

In model NoE, the filling factor of the DRs is simply assumed to be unity.
Namely, the turbulence is distributed over the whole FBs.
In this case, there is no effect of electron escape.
Once electrons are injected, they are continuously affected by the turbulence.

%%%%%%%%%%%%%%%%%%%%%%%
\begin{figure}[!htb]
\begin{center}
\epsscale{1.0}
\plotone{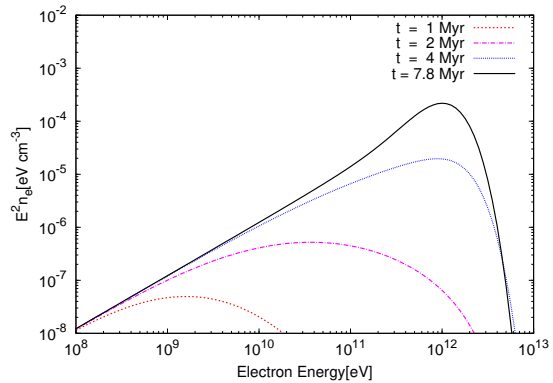}
\caption{Evolution of the electron spectrum for the NoE model.}
\label{ele3}
\end{center}
\end{figure}
%%%%%%%%%%%%%%%%%%%%%%%

In Figure \ref{ele3}, we plot the evolution of the electron spectrum.
Each time in the figure (1, 2, 4, 7.8 Myr) corresponds to the distance from
the shock front (0.26, 0.51, 1.0, 2.0 kpc, respectively).
Apparently, the maximum energy of electrons gradually grows with time.
The energy distribution is not the steady-state solution,
since the initial acceleration timescale
$p^2/D_{pp}(0)= 3/(2 \beta_{\rm W}^2 D_0)\simeq 1.3$ Myr,
which is independent of electron energy, and
the cooling timescale
%%%%%%%%%%%%%%%%%%%%%%%
\begin{eqnarray}
t_{\rm cool} &=& \frac{3 m_{\rm e} c}{4 \sigma_{\rm T} \gamma U_{\rm ph}} \nonumber \\
&\simeq& 340 \left( \frac{E}{10^9~\mbox{eV}} \right)^{-1}
\left( \frac{U_{\rm ph}}{1.5 \times 10^{-12}~\mbox{erg}~\mbox{cm}^{-3}} \right)^{-1}
\mbox{Myr}, \nonumber \\
\end{eqnarray}
%%%%%%%%%%%%%%%%%%%%%%%
are long enough compared to the elapsed time.
The electron spectrum becomes harder with increasing time.
The cooling effect is not so sufficient that the energy density
of electrons increases with time or $x$.
At later stages (see the line of 7.8 Myr),
the spectrum becomes similar to the steady solution,
in which a spectral bump at $\sim 10^{12}$ eV
is formed by the balance of the acceleration
and cooling.

As we have mentioned, even if we adopt the same model as MS11
with the escape effect, the time-dependent calculation
yields a similar evolution of $n$ to that in Figure \ref{ele3}.
This is an opposite manner of the spatial dependence of
the electron spectrum in MS11;
the steady solution in MS11 shows a hard-to-soft distribution
from the shock front to the downstream.
Note that the escape timescale in MS11 is $\sim 100$--$400$ Myr
so that the escape effect can be neglected.
The steady-state solutions may not be appropriate to
describe the electron energy distribution.

%%%%%%%%%%%%%%%%%%%%%%%
\begin{figure*}[!htb]
\begin{center}
\epsscale{1.1}
\plottwo{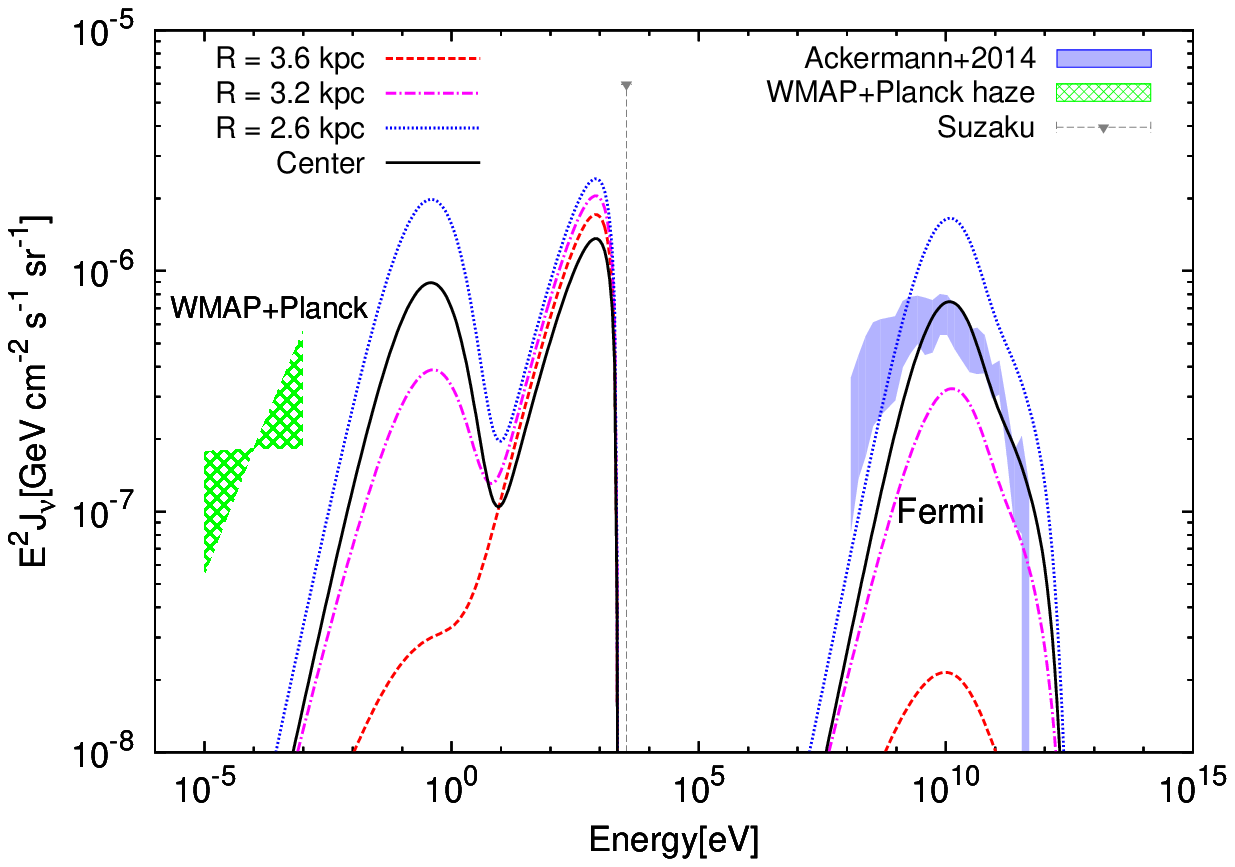}{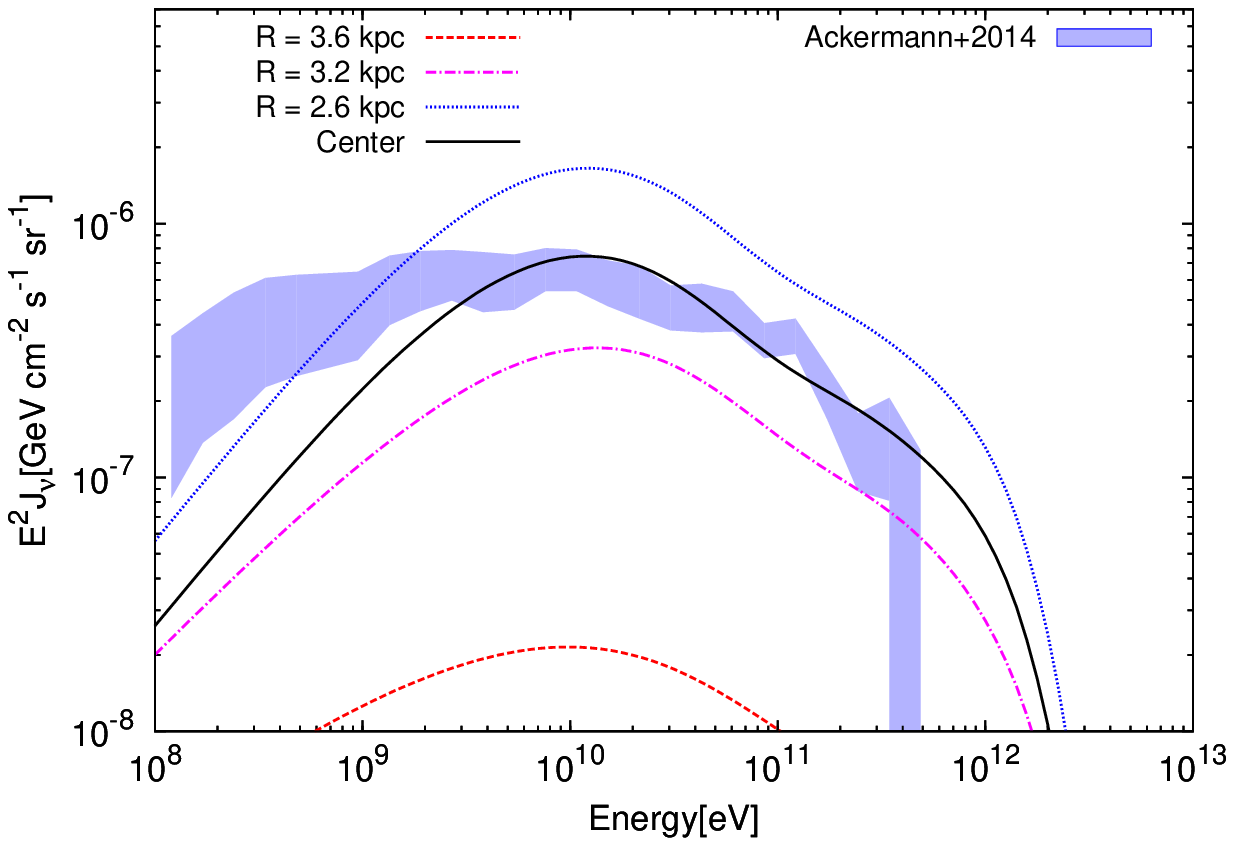}
\caption{Broadband photon spectra of the NoE model (left)
and their zoom-in display in the gamma-ray band (right).
The thermal bremsstrahlung spectrum in the X-ray band
is plotted assuming a proton density of $10^{-3}~\mbox{cm}^{-3}$
and electron temperature of keV for reference.
The observed data are taken from \citet{ack14} for the Fermi-LAT data,
and \citet{kat13} for the WMAP+Planck data and the X-ray upper limit.}
\label{spe3}
\end{center}
\end{figure*}
%%%%%%%%%%%%%%%%%%%%%%%

%%%%%%%%%%%%%%%%%%%%%%%
\begin{figure*}[!htb]
\begin{center}
\epsscale{1.1}
\plottwo{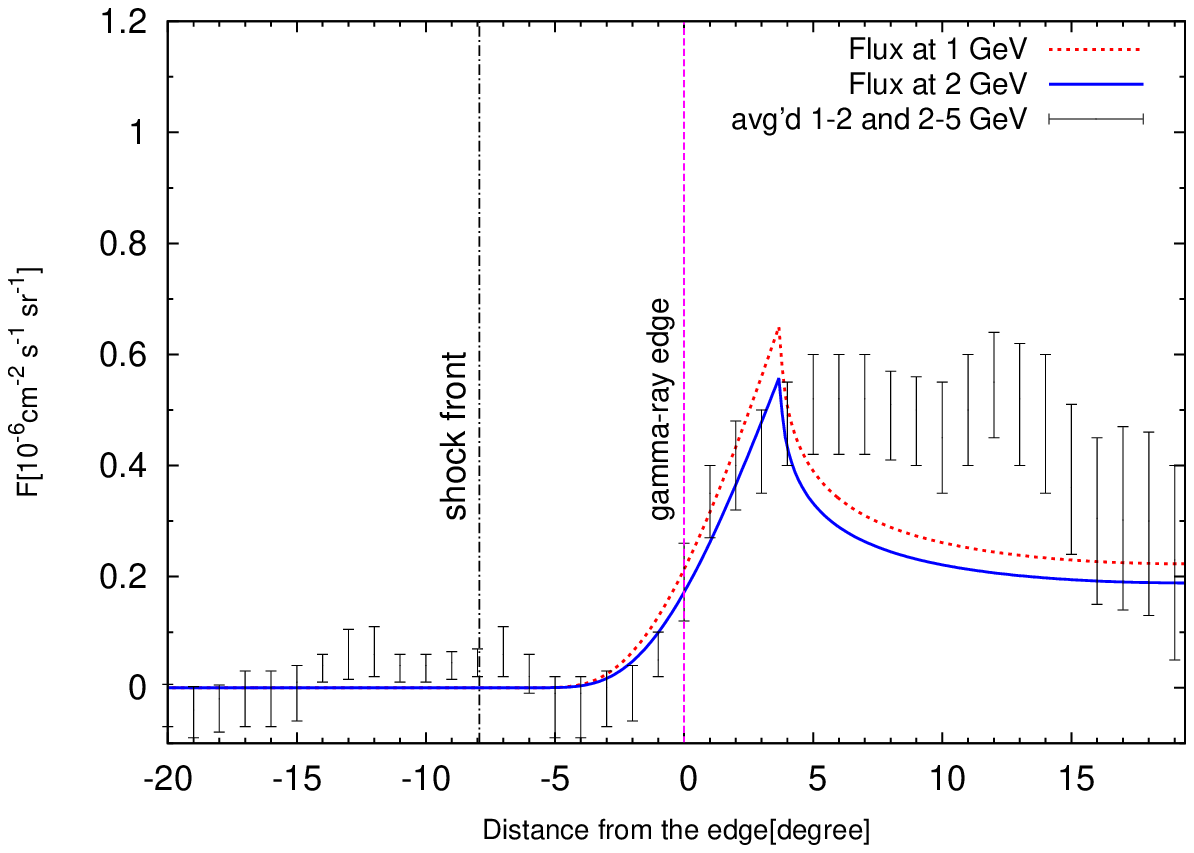}{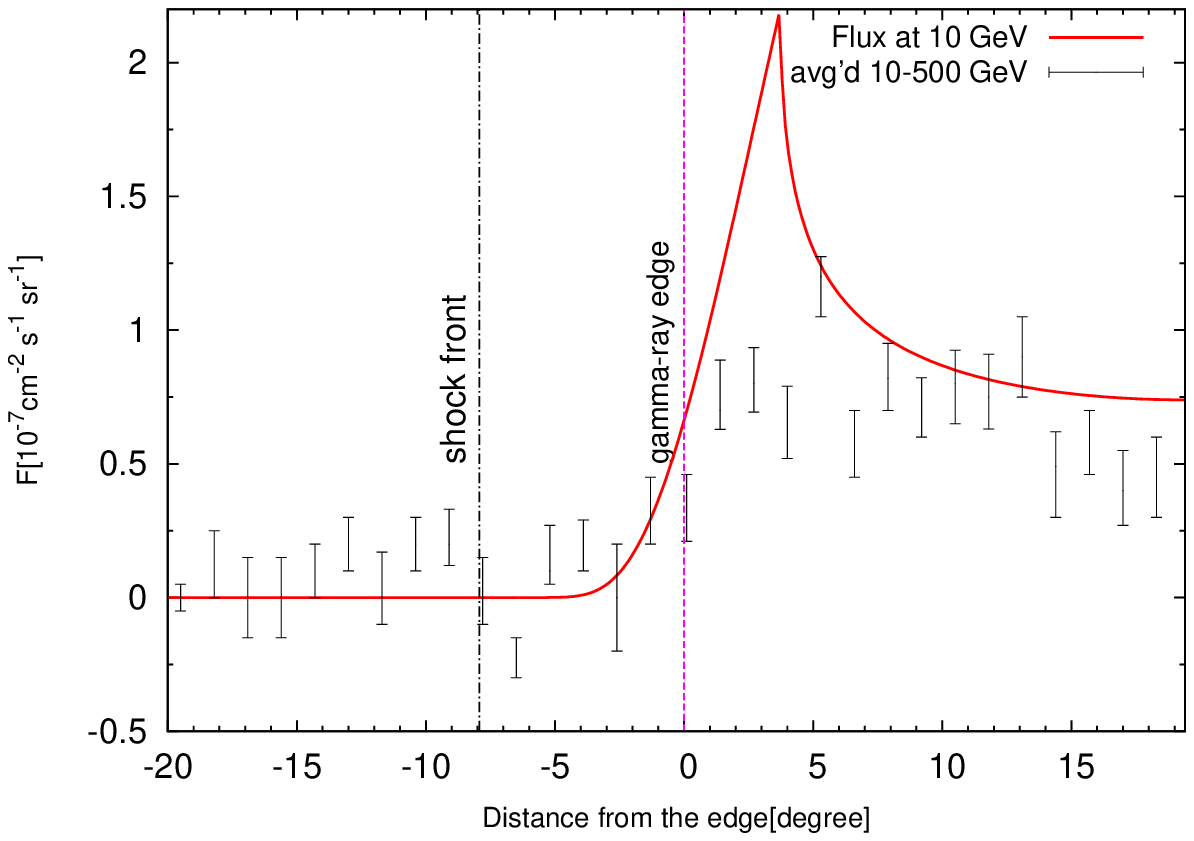}
\caption{Gamma-ray intensity of the NoE model
as a function of distance from the gamma-ray edge
for 1 GeV (dotted) and 2 GeV (solid) profiles (left panel)
with the data from \citet{su10},
and for 10 GeV (solid) profile (right panel) with the data from \citet{ack14}.}
\label{sur3}
\end{center}
\end{figure*}
%%%%%%%%%%%%%%%%%%%%%%%

In our model, the electron spectrum is maintained hard
during the calculation, because the escape effect is neglected.
As a result, the gamma-ray spectrum below 1 GeV becomes too hard,
as shown in Figure \ref{spe3}.
Another problem is that the surface brightness monotonically
increases as the radius approaches to the center.
As shown in Figure \ref{sur3}, the inner boundary at $x=2$ kpc
is reflected on the surface brightness as a sharp peak.
Unless such an artificial cavity exists,
the surface brightness profile should have a peak at the center
in this model.
The nearly homogeneous surface brightness is difficult to be reproduced
with this model.

The photon spectra and surface brightness
obtained by the time-dependent calculation with the same model as MS11
are also similar to Figures \ref{spe3} and \ref{sur3}.
In order to soften the spectrum, especially for the inner region,
the escape effect should be incorporated.

\subsection{Mild Escape Model}

\begin{figure}[!htb]
\begin{center}
\epsscale{1.0}
\plotone{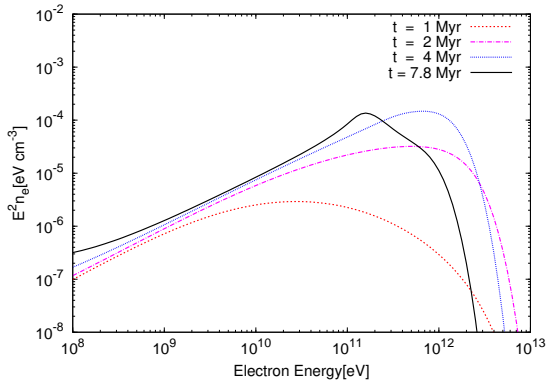}
\caption{Evolution of the total electron spectrum for the MiE model.}
\label{ele1}
\end{center}
\end{figure}

To soften the electron and photon spectra,
the escape effect should be involved.
In the MiE model, we take into account the escape effect
with the initial timescale of $t_{\rm esc}=3 L^2 D_0/c^2
=13$ Myr, which is much longer than $t_0=L/v_{\rm sh}=0.7$ Myr
and the acceleration timescale, 0.48 Myr.
As time passes, the escape timescale shortens,
while the acceleration timescale grows.
In Figure \ref{ele1}, we show the evolution of the total
electron spectrum (namely, $n+n_{\rm esc}$).
At $t=2$ Myr, the escape effect is still negligible,
so the electrons in the DRs still dominate for the entire energy region.
Even at $t=4$ Myr, the peak energy region ($\sim 10^{12}$ eV)
is dominated by the electrons in the acceleration process.
However, in the low-energy power-law-like region (below $\sim 10^{11}$ eV),
the escaped electron density is comparable to that in the DRs.
At the last stage ($t=7.8$ Myr),
the IC cooling makes a spectral peak for the escaped electrons
at $\sim 10^{11}$ eV, while the highest energy region ($\sim 10^{12}$ eV)
is still dominated by the electrons in the DRs.

%%%%%%%%%%%%%%%%%%%%%%%
%%%%%%%%%%%%%%%%%%%%%%%

\begin{figure*}[!htb]
\begin{center}
\epsscale{1.1}
\plottwo{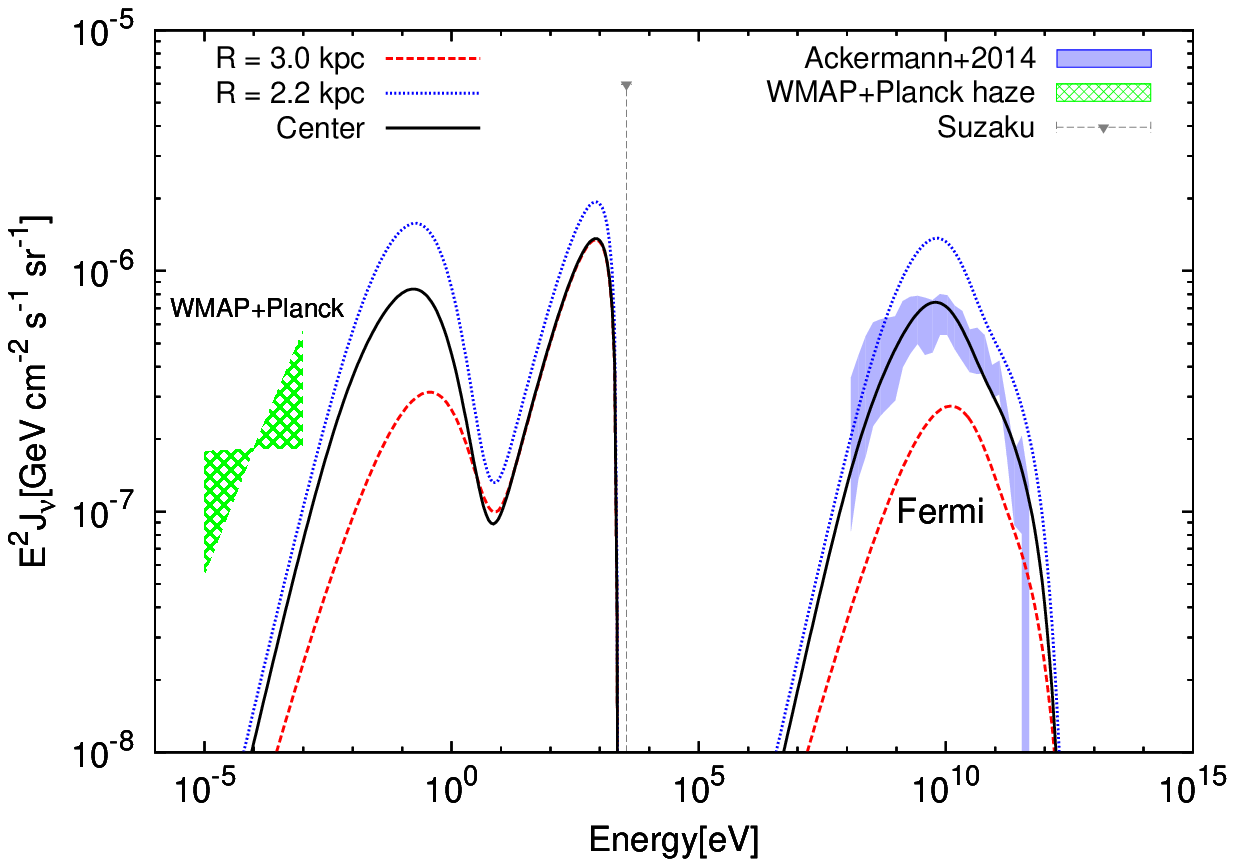}{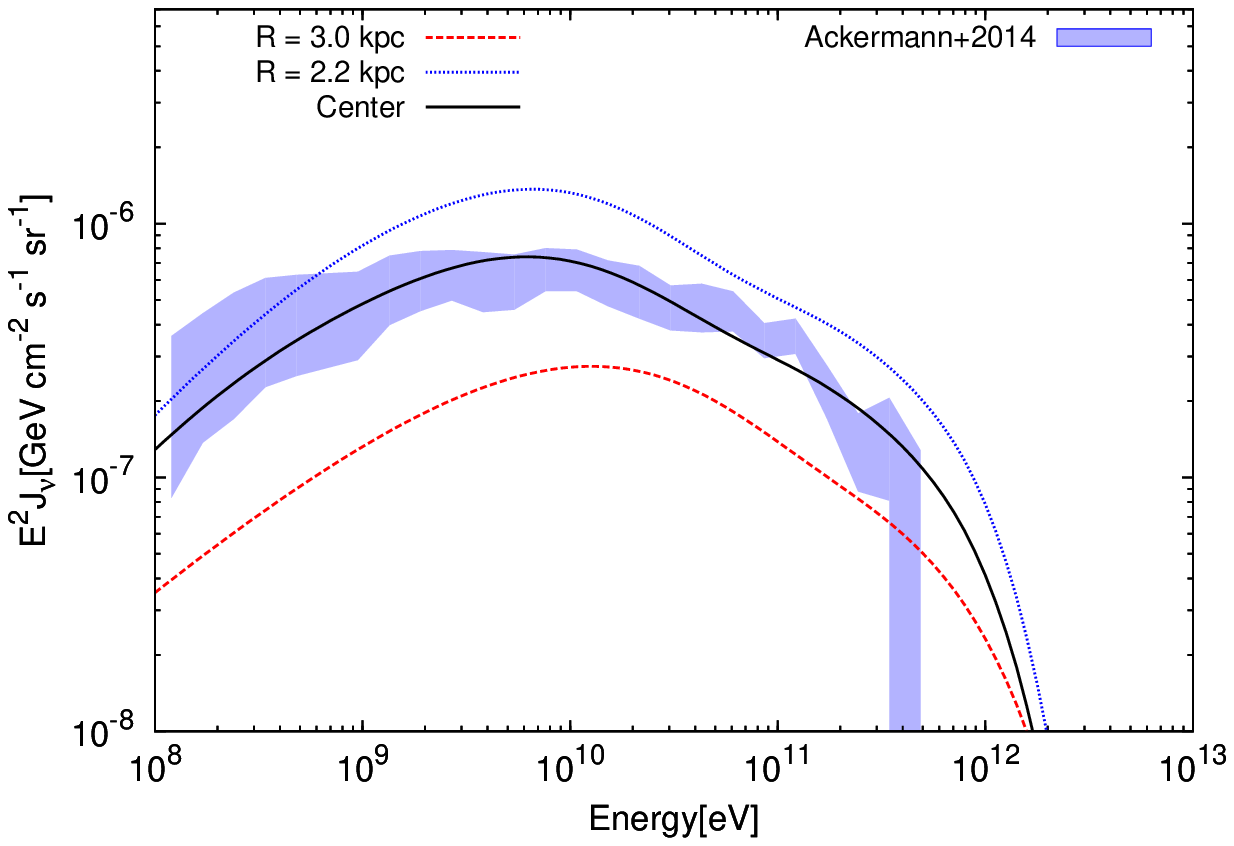}
\caption{Same as Fig. \ref{spe3} but for the MiE model.}
\label{spe1}
\end{center}
\end{figure*}

\begin{figure*}[!htb]
\begin{center}
\epsscale{1.1}
\plottwo{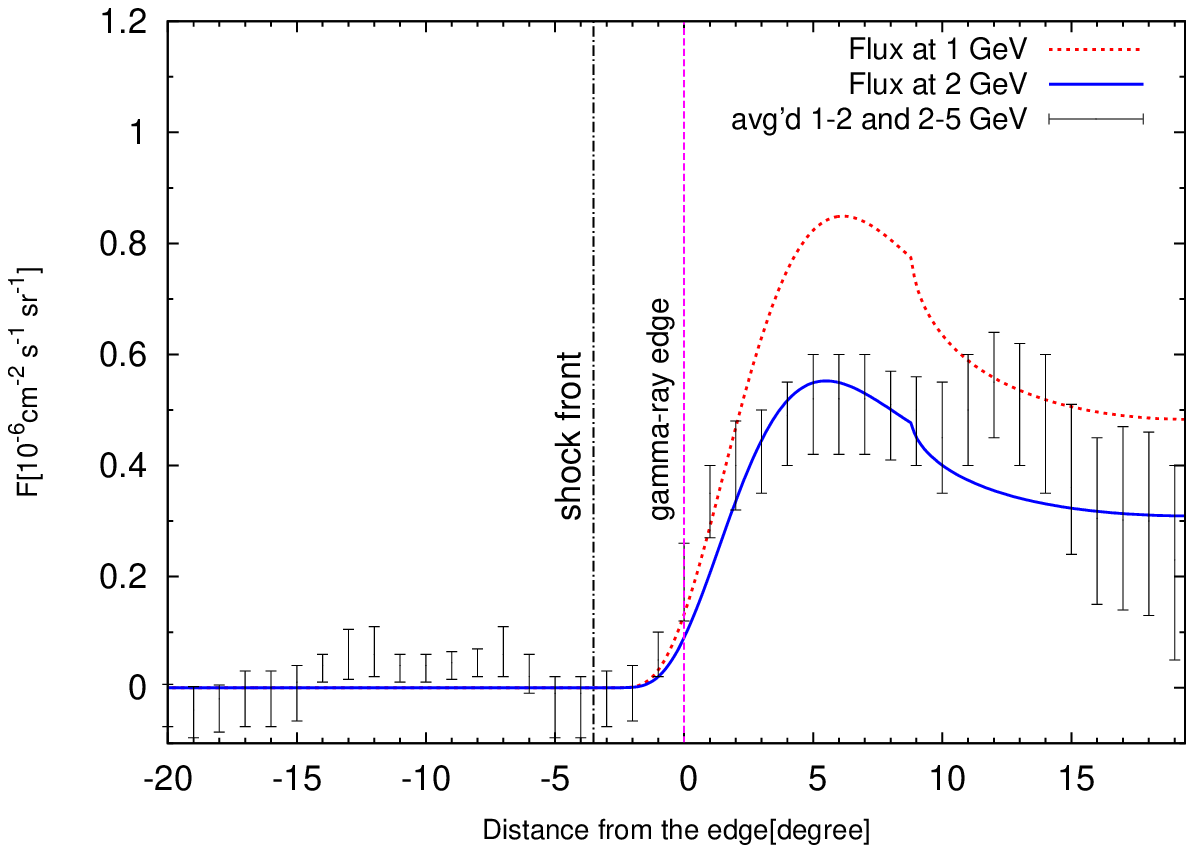}{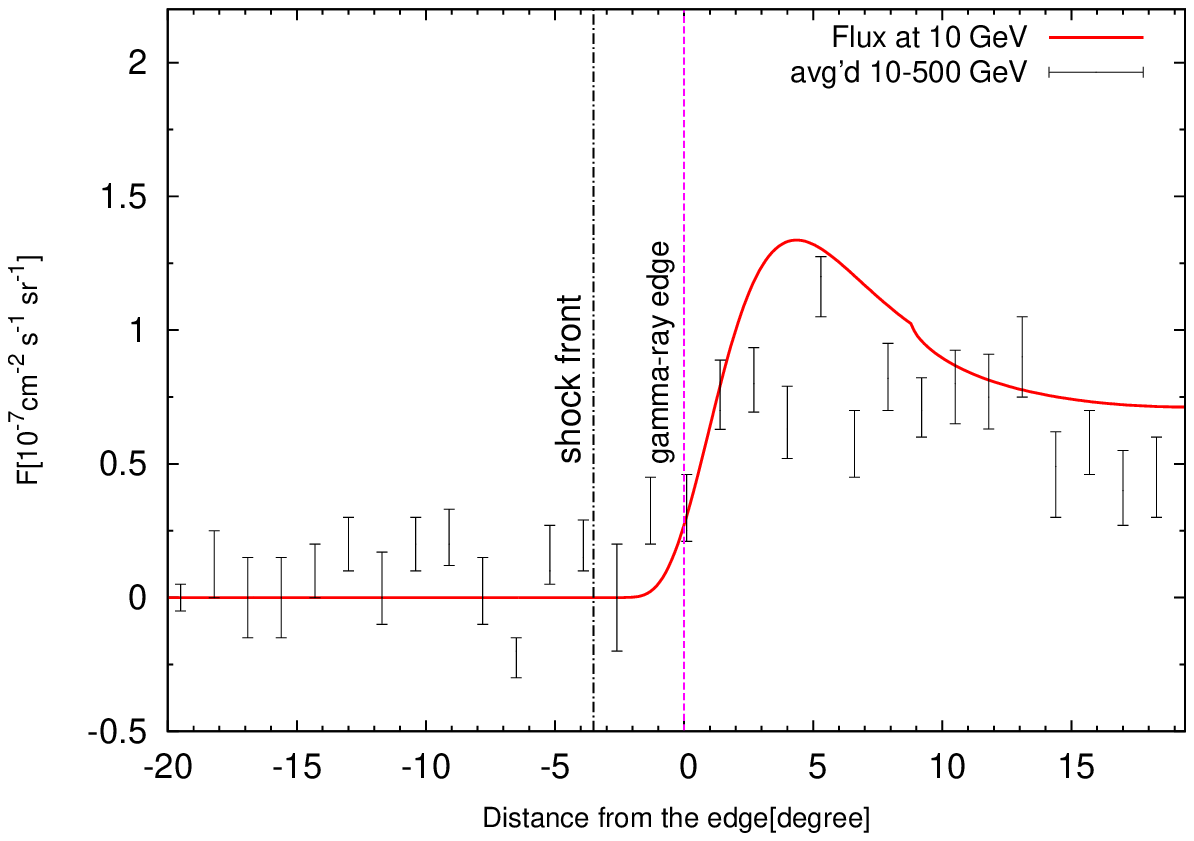}
\caption{Same as Fig. \ref{sur3} but for the MiE model.}
\label{sur1}
\end{center}
\end{figure*}

The softening of the electron spectrum due to the IC cooling and escape effects
understandably produces a softer gamma-ray spectrum than the NoE model spectrum.
As shown in Figure \ref{spe1}, the resultant spectrum agrees well with
the observation above $10^8$ eV.
The surface brightness profiles also agree with the observed
properties: sharp rise at the edge and flat profile inside
(see Figure \ref{sur1}).

%%%%%%%%%%%%%%%%%%%%%%%
%%%%%%%%%%%%%%%%%%%%%%%

%%%%%%%%%%%%%%%%%%%%%%%
%%%%%%%%%%%%%%%%%%%%%%%

Even with the best-fit parameters for the MiE model,
the radio intensity is far below the observed data of the WMAP haze
(see Figure \ref{spe1}).
While we have assumed $B_0=4 \mu\mbox{G}$,
a stronger magnetic field such as 10--20 $\mu$G may be required as shown in MS11
\citep[see also, e.g.][]{ack14} to reconcile the radio intensity.
However, the magnetic field model in \citet{orl13},
which agrees with all-sky total intensity and polarization maps,
may not allow such a high field.

\subsection{Efficient Escape Model}

In this subsection, we try to find a model
that agrees with not only the gamma-ray spectrum
but also the WMAP haze.
The electron spectrum should be softer than
that in the MiE model.
We adopt a very small size for $L$ in the EfE model,
so the escape effect will be more prominent.
To accelerate electrons as far as $10^{12}$ eV before
escaping from the DRs, a higher value for $D_0$
is adopted here.
 
%%%%%%%%%%%%%%%%%%%%%%%
\begin{figure}[!htb]
\begin{center}
\epsscale{1.0}
\plotone{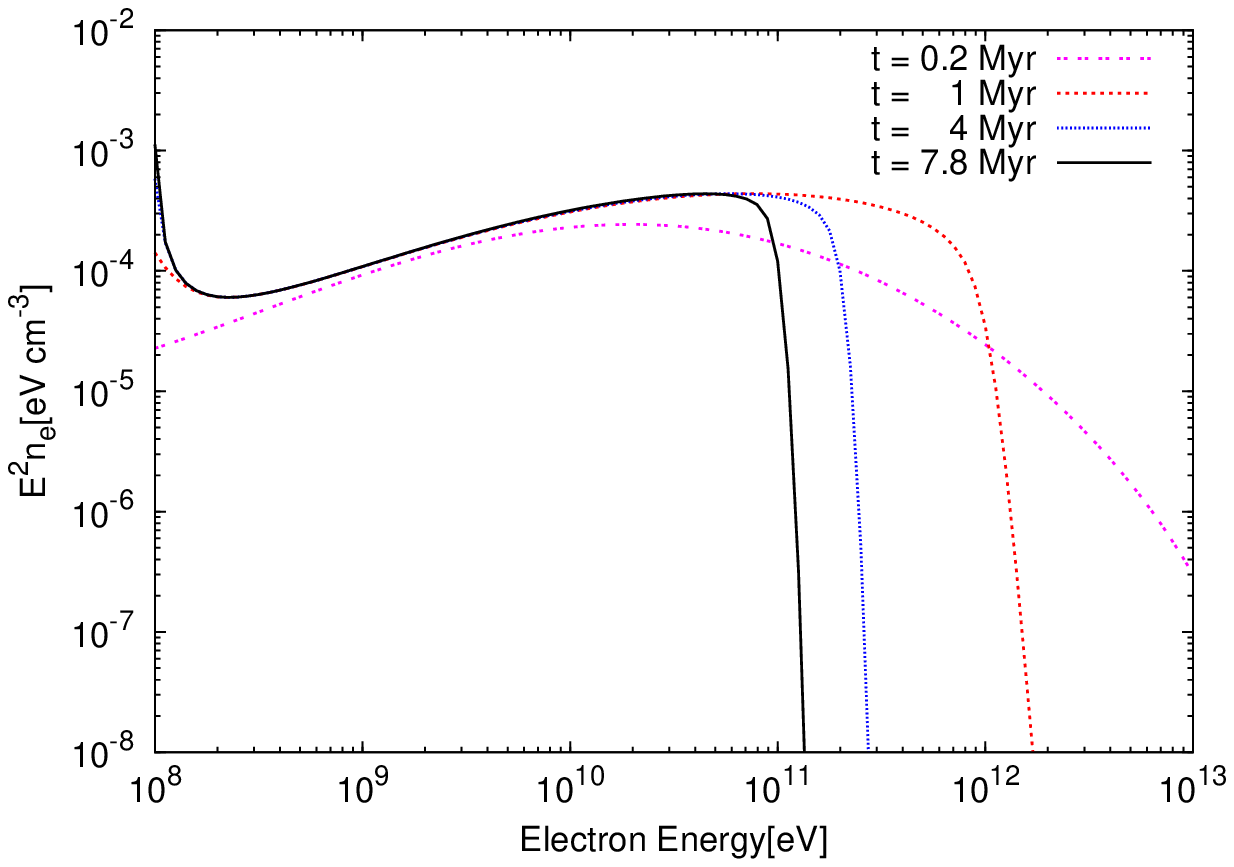}
\caption{Evolution of the total electron spectrum for the EfE model.}
\label{ele2}
\end{center}
\end{figure}
%%%%%%%%%%%%%%%%%%%%%%%

The initial timescales are $t_0=0.039$ Myr,
$t_{\rm acc}=0.048$ Myr, and $t_{\rm esc}=0.40$ Myr.
At $t=0.086$ Myr, the acceleration and escape timescales
become comparable as $\sim 0.13$ Myr.
At $t=0.2$ Myr (see Figure \ref{ele2}), most electrons
above $10^{10}$ eV are still in the DRs.
However, the escaped electrons become dominant
in the later phase ($t>0.3$ Myr).
Then, the acceleration efficiency quickly damps.
The inefficiency of the acceleration in the later phase is due to not only
the escape effect but also the growth of the acceleration
timescale.
The non-steady electron spectrum in the early acceleration stage
is therefore frozen after the decay of the acceleration efficiency.
This leads to the soft electron spectrum.
Owing to the fast decay of the acceleration efficiency,
the electrons injected later almost remain at the injection energy
so that an artificial spectral peak is seen at $10^8$ eV.
The spectral cutoff due to the IC cooling
($t_{\rm cool}=0.34 (E/10^{12}~\mbox{eV})^{-1}$ Myr)
is clearly seen in the electron spectra for $t \geq 1$ Myr
in Figure \ref{ele2}.

%%%%%%%%%%%%%%%%%%%%%%%
\begin{figure*}[!htb]
\begin{center}
\epsscale{1.1}
\plottwo{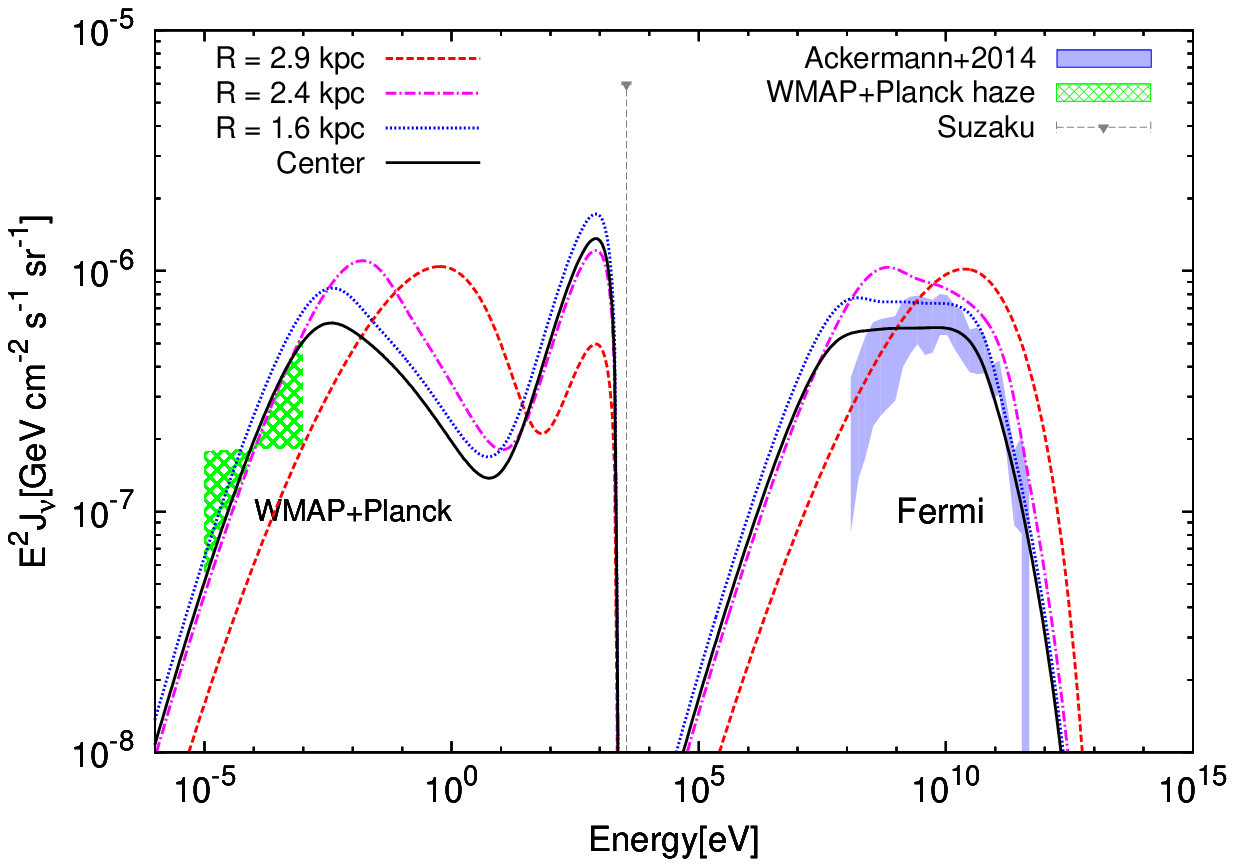}{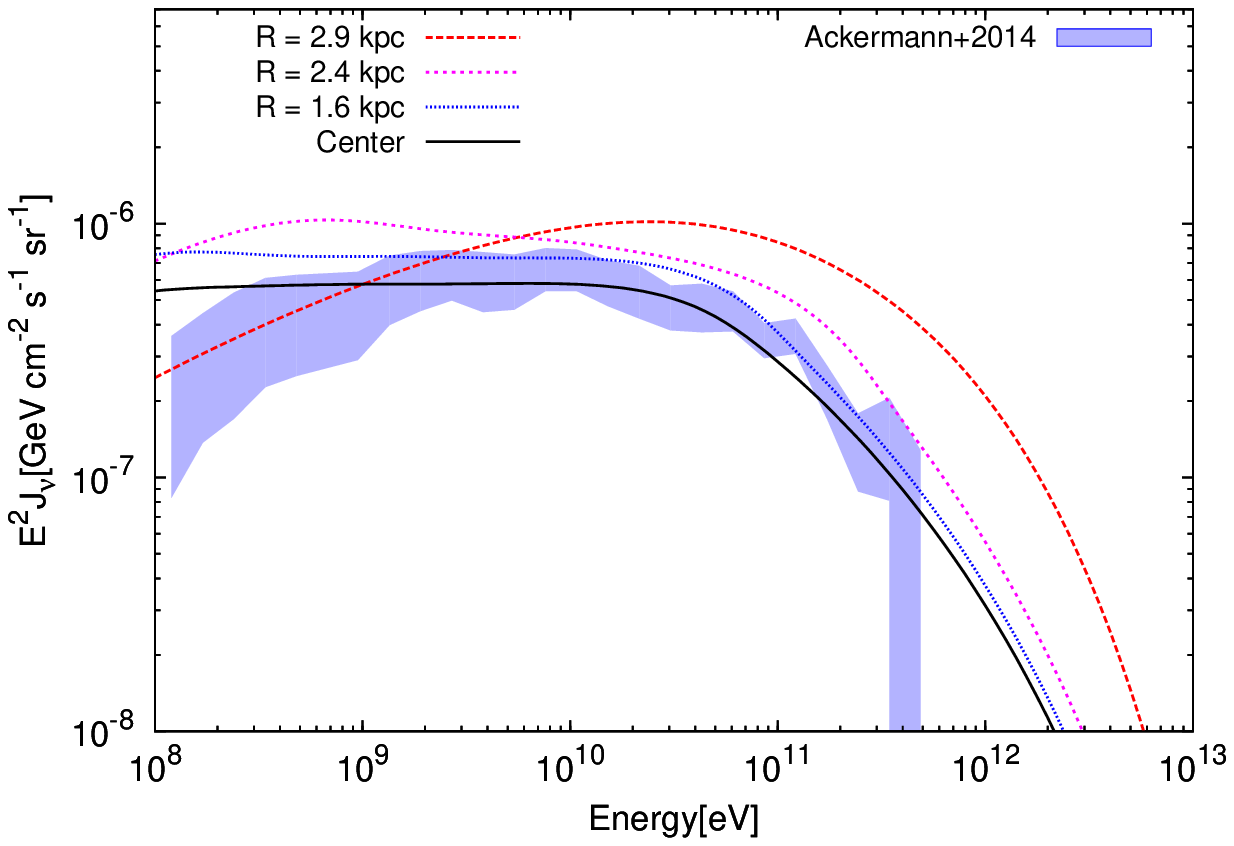}
\caption{Same as Fig. \ref{spe3} but for the EfE model.}
\label{spe2}
\end{center}
\end{figure*}
%%%%%%%%%%%%%%%%%%%%%%%

%%%%%%%%%%%%%%%%%%%%%%%
\begin{figure*}[!htb]
\begin{center}
\epsscale{1.1}
\plottwo{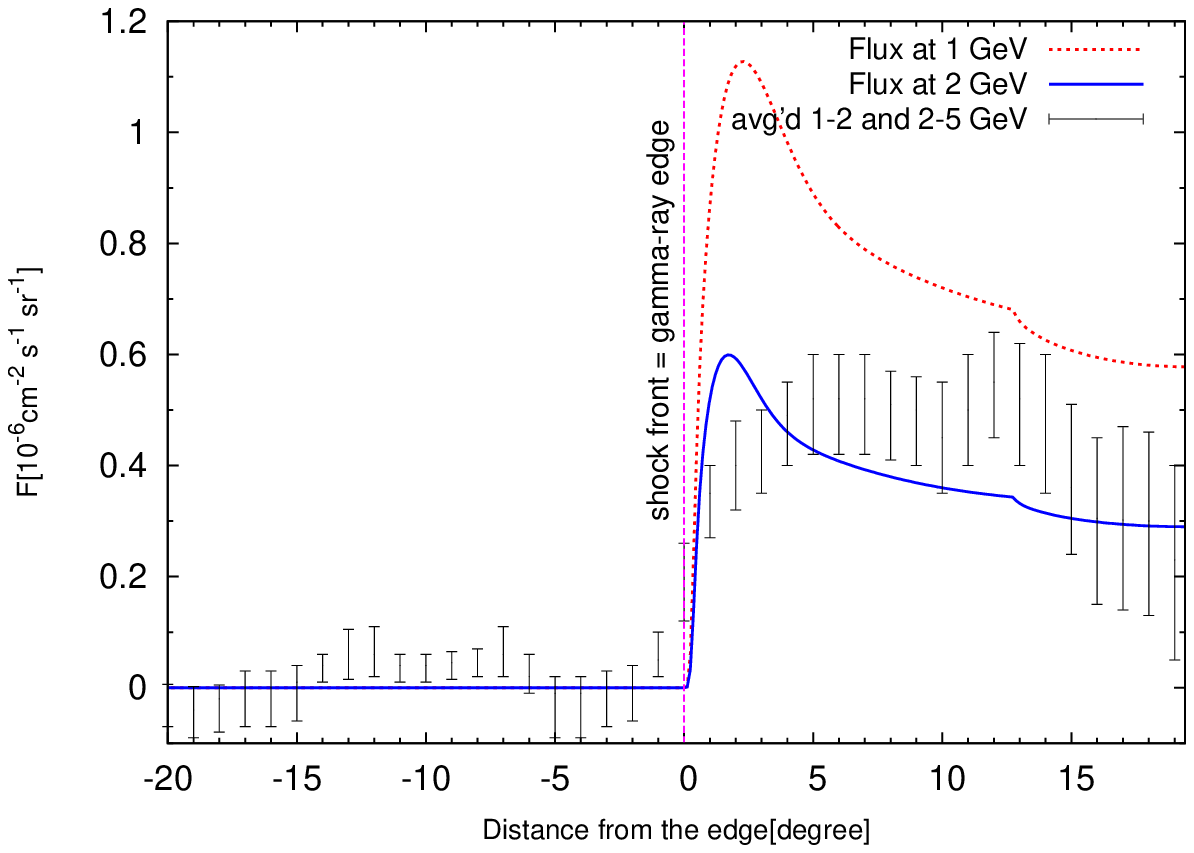}{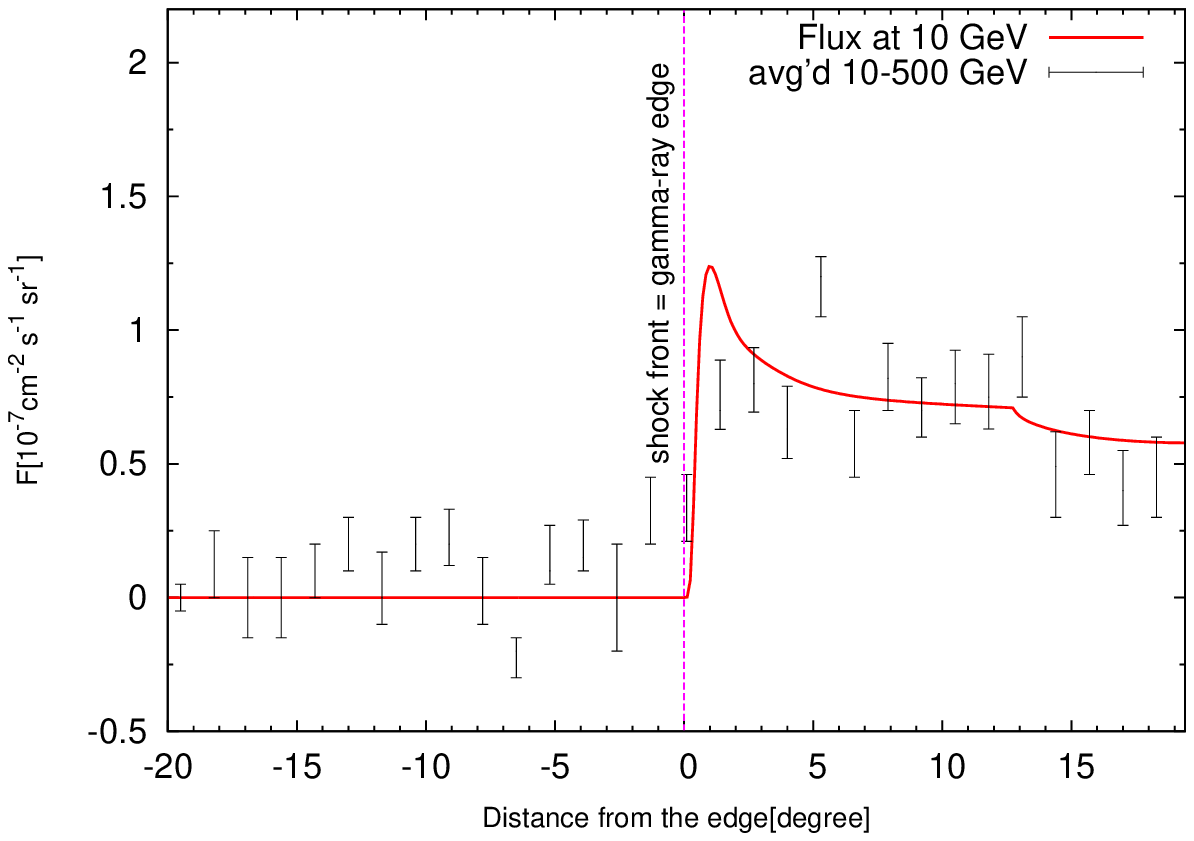}
\caption{Same as Fig. \ref{sur3} but for the EfE model.}
\label{sur2}
\end{center}
\end{figure*}
%%%%%%%%%%%%%%%%%%%%%%%

The resultant photon spectrum becomes so soft that
the radio intensity of the WMAP haze is also reproduced
as seen in Figure \ref{spe2}.
The model flux at $10^8$ eV is slightly higher than the observation,
but may still be within the systematic uncertainty.
The surface brightness profiles do not seem to deviate
from the observed profiles very much (see Figure \ref{sur2}),
but the short acceleration period makes
a slight limb brightening.
In this model, the initial acceleration timescale is much shorter
than the previous models.
Therefore, the positions of the shock front and the FB edge
are identical in this case.

\section{Hadronic Models}
\label{sec:had}

%%%%%%%%%%%%%%%%%%%%%%%
\begin{deluxetable*}{lccccc}
\tablewidth{0pt}
\tablecaption{Parameters for hadronic models \label{para_had}}
\tablehead{
\colhead{Model}                    & \colhead{$L$ (pc)}                               &
\colhead{$D_0$ ($\mbox{s}^{-1}$)}  & \colhead{$Q_0$ ($\mbox{cm}^{-3}~\mbox{s}^{-1}$)} &
\colhead{$R_{\rm sh}$ (kpc)}       & \colhead{$E_{\rm tot}$ (erg)} 
}
\startdata
pNoE (No Escape) & 1500\tablenotemark{a} & $1.0 \times 10^{-7}$ &
$1.1 \times 10^{-25}$ & 4.4 & $1.4 \times 10^{57}$ \\
pMiE (Mild Escape) & 180 & $1.4 \times 10^{-7}$ &
$1.8 \times 10^{-25}$ & 3.6 & $ 1.4 \times10^{57}$ \\
pEfE (Efficient Escape) & 10 & $1.2 \times 10^{-6}$ &
$3.9 \times10^{-25}$ & 3.0 & $2.2 \times 10^{57}$ \\
\enddata
\tablenotetext{a}{In model pNoE, we neglect the escape effect,
so $L$ provides the timescale $t_0$ only.}
\end{deluxetable*}
%%%%%%%%%%%%%%%%%%%%%%%

%%%%%%%%%%%%%%%%%%%%%%%
\begin{figure}[!htb]
\begin{center}
\epsscale{1.0}
\plotone{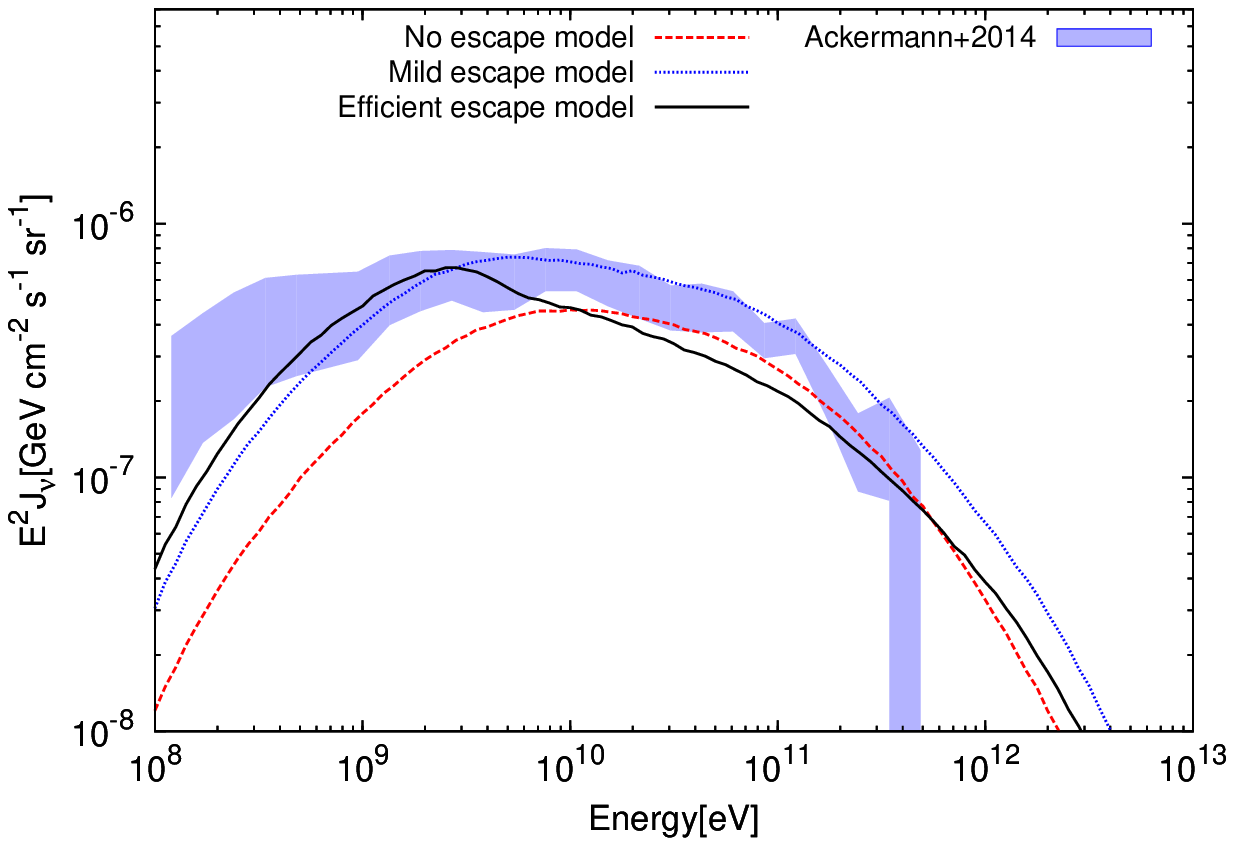}
\caption{Photon spectra at the center of the FBs
for three hadronic models.
The observed data are taken from \citet{ack14}.}
\label{gspe_pro}
\end{center}
\end{figure}
%%%%%%%%%%%%%%%%%%%%%%%

A hadronic model with stochastic acceleration
is also possible in theory.
The acceleration mechanism is common for electrons and protons
other than the injection process.
The accelerated protons should obviously exist in the situation
we have considered in this paper.
High-energy protons collide with the thermal protons,
and produce pions that decay into gamma-rays,
which may be responsible for the gamma-ray emission in the FBs.
Similarly to the leptonic models, we test three models
as summarized in Table \ref{para_had}.
The radiative cooling effect is neglected so that
the difference between protons in the DRs and escaped protons
is the acceleration process only.
We adopt $\alpha=1$ again.
The background density of protons as the targets for $pp$-collision
is assumed to be uniform with $10^{-3}~\mbox{cm}^{-3}$.

In Figure \ref{gspe_pro}, we plot the gamma-ray spectra
at the center ($l=0^\circ$) for the three models.
As the escape effect becomes efficient,
the photon spectrum softens.
Although the hadronic models can generate a photon spectrum
similar to the observed one, the required total energy
becomes much larger than those in the leptonic models
(compare $E_{\rm tot}$ in Tables \ref{para_lep} and \ref{para_had},
where total energies integrated over the whole volume are shown).
In such cases, electrons should be accelerated as well.
Even if the electron energy fraction to the proton energy
is plausibly theoretical minimum-value, $m_{\rm e}/m_{\rm p}$,
the gamma-ray emission would be dominated by IC emission by electrons.
Therefore, the hadronic models with the stochastic acceleration
seem contrived.
Furthermore,
the synchrotron flux from secondary electrons/positrons
is too dim to reconcile the WMAP haze.

\section{Summary and Discussion}
\label{sec:sum}

In our model, the shock fronts are propagating outward with
a low Mach number so that the direct particle acceleration
by the shock waves may be inefficient.
Some kind of turbulence is induced just behind the shock front,
and gradually decays with time.
Particles are accelerated via scattering with the turbulence.
Considering the efficiency of gamma-ray emission,
the leptonic model is more likely than the hadronic model
for stochastic acceleration models.
The adequate acceleration timescale
is initially $0.05$--$0.5$ Myr.
Our results show that the steady solutions
for the Fokker--Planck equation are not adequate in the FBs.
The time interval in which the maximum electron energy reaches
$10^{12}$ eV is not negligible compared to the bubble
expansion timescale.
We also show that the escape effect may be indispensable
to reproduce the gamma-ray spectrum especially
below 1 GeV.
If electrons are continuously accelerated,
the electron spectrum becomes too hard to agree with the observed gamma-ray spectrum.
For the escaped electrons,
the termination of the acceleration and succeeding cooling effect
generate a soft spectrum.
The combination of the two electron species,
those in the acceleration region and those having escaped,
can produce the best-fit model for the FBs.
In addition, the long-term evolution of the electron spectrum,
which is regulated by the finite timescales of $t_0$, $t_{\rm acc}$,
$t_{\rm esc}$, and $t_{\rm cool}$,
is favorable for the almost constant surface brightness.
Moreover, assuming a very short $t_{\rm esc}$,
we can reproduce the spectra of both the FBs and WMAP haze
with the reasonable magnetic field.

While we cannot determine the model parameters uniquely,
our results modestly constrain a parameter range.
If the turbulence decay is much faster than our models,
the initial acceleration efficiency should be high.
In such cases, the particle acceleration and gamma-ray emission
occur near the shock front only.
This leads to significant limb brightening in the surface brightness profile
as seen in the synchrotron emission
from shock-accelerated electrons in supernova remnants.
So a too small $t_0$ (or equivalently $L$) or $\alpha=2$
seem unfavorable.

In order to make electrons escape from the acceleration process
in a certain timescale, we assume a finite size of the DR.
An inhomogeneity in the upstream region is required
with a significant filling factor.
The region size $L$ required in our model is comparable to
the size of the inhomogeneity expected from several studies,
such as the numerical simulations of
the interaction between a galactic wind and hot halo gas \citep{mel13,sha14}
or formation of galactic fountain clouds \citep{fra15}.
This coincidence encourages us assuming such
localized acceleration regions.

The average electron density at $x=2$ kpc obtained from Table \ref{para_lep}
is $2.3 \times 10^{-15}~\mbox{cm}^{-3}$ in the MiE model.
Those electrons may be injected from the background plasma.
Though the cosmic-ray electron density in the halo is highly uncertain
\citep[e.g.][]{orl13},
that can be 1--50\% of the local density, $\sim 10^{-13}~\mbox{cm}^{-3}$
\citep{ack10}.
Considering the uncertainty of the filling factor,
the re-acceleration of the background cosmic-ray electrons
seems a promising model for the electron injection mechanism.

The spatial diffusion coefficient inside the Galactic disk
is $5 \times 10^{28}~\mbox{cm}^2~\mbox{s}^{-1}$ at $3$--$4$ GeV
\citep{ack12}.
If a Kolmogorov-like turbulence dominates ($D_{xx} \propto p^{1/3}$),
the diffusion coefficient at $10^{12}$ eV becomes $\sim
10^{29}~\mbox{cm}^2~\mbox{s}^{-1}$.
On the other hand, our model parameters infer
$D_{xx}=7.5 \times 10^{25}$--$7.5 \times 10^{26}~\mbox{cm}^2~\mbox{s}^{-1}$
independently of the particles' energy.
As a matter of course, the mean free path in the DRs
($10^{-3}$--$10^{-2}$ pc) is much shorter than
the local value near the Sun ($\sim$ pc at $10^{12}$ eV),
but much longer than the Larmor radius.

\acknowledgements

We thank the anonymous referee for a careful review
and valuable comments.
We appreciate Y. Fujita and Y. Ohira for helpful discussion.
This work is partially supported by the Grant-in-Aid for
Scientific Research, No. 25400227 from the MEXT of Japan (KA).

\end{document}